\documentclass[10pt,journal,compsoc]{IEEEtran}
\ifCLASSOPTIONcompsoc
  \usepackage[nocompress]{cite}
\else
  \usepackage{cite}
\fi
\ifCLASSINFOpdf
   \usepackage[pdftex]{graphicx}
  \graphicspath{{../figures/}}
   \DeclareGraphicsExtensions{.pdf,.jpeg,.png, .svg}
\else
\fi
\usepackage{svg}

\usepackage{amsmath}
\usepackage{algpseudocode}

\usepackage{url}

\usepackage[ruled, linesnumbered]{algorithm2e}
\usepackage{amsmath}
\DeclareMathOperator*{\argmin}{argmin} 
\usepackage{xcolor}
\usepackage{wrapfig}
\begin{document}
%
\title{Submerse: Visualizing Storm Surge Flooding Simulations in Immersive Display Ecologies}
%
%
%
%

\author{Saeed~Boorboor,
        Yoonsang~Kim,
        Ping~Hu,
        Josef~M.~Moses,
        Brian~A.~Colle,
        and~Arie~E.~Kaufman,~\IEEEmembership{Fellow,~IEEE}
\IEEEcompsocitemizethanks{\IEEEcompsocthanksitem Boorboor, Kim, Hu, and Kaufman are with the Department
of Computer Science, Stony Brook University.\protect\\
E-mail: \{sboorboor, yoonsakim, pihu, ari\}@cs.stonybrook.edu.
\IEEEcompsocthanksitem  Moses and Colle are with the School of Marine and Atmospheric Sciences, Stony Brook University. }
\thanks{Manuscript received xxxxx; revised xxxxx.}}

%
%

\markboth{Journal of \LaTeX\ Class Files,~Vol.~14, No.~8, August~2015}%
{Boorboor \MakeLowercase{\textit{et al.}}: Submerse}
%



\IEEEtitleabstractindextext{%
\begin{abstract}
We present \textit{Submerse}, an end-to-end framework for visualizing flooding scenarios on large and immersive display ecologies. 
Specifically, we reconstruct a surface mesh from input flood simulation data and generate a to-scale 3D virtual scene by incorporating geographical data such as terrain, textures, buildings, and additional scene objects.
To optimize computation and memory performance for large simulation datasets, we discretize the data on an adaptive grid using dynamic quadtrees and support level-of-detail based rendering.  
Moreover, to provide a perception of flooding direction for a time instance, we animate the surface mesh by synthesizing water waves.
As interaction is key for effective decision-making and analysis, we introduce two novel techniques for flood visualization in immersive systems: (1) an automatic scene-navigation method using optimal camera viewpoints generated for marked points-of-interest based on the display layout, and (2) an AR-based focus+context technique using an aux display system.
Submerse is developed in collaboration between computer scientists and atmospheric scientists.
We evaluate the effectiveness of our system and application by conducting workshops with emergency managers, domain experts, and concerned stakeholders in the Stony Brook Reality Deck, an immersive gigapixel facility, to visualize a superstorm flooding scenario in New~York~City. 

\end{abstract}

}

\maketitle

\IEEEdisplaynontitleabstractindextext

%
\IEEEpeerreviewmaketitle

\IEEEraisesectionheading{\section{Introduction}\label{sec:introduction}}

%
%
%
%
\IEEEPARstart{T}{he} increase in extreme weather events~\cite{field2014summary} has led to a growing need for visualization systems that can aid scientists, emergency managers, and concerned stakeholders to effectively prepare for the impacts of potential catastrophic events.
Of particular concern is coastal and inland flooding.
Studies have shown that floods account for 43\% of all disasters and are the most frequently occurring natural disasters worldwide~\cite{the_international_disasters_database}.
Although there have been advances in weather forecasting models, their usefulness depends greatly on the human decision-making dimension~\cite{murphy1993good}. 

Predicted flooding scenarios using numerical simulation models are typically rendered as 2D visualizations superimposed onto 2D maps.
While a 2D representation can convey the birds-eye view extent of the flooding, it relies on the user to perceive depth information based on cues such as color.
The use of 3D graphics and virtual reality (VR) has been shown to positively influence perceptions and behaviors in preparing for and responding to natural hazards~\cite{mol2022after}.
Extending the visualizations to 3D~\cite{cornel2015visualization, cornel2019interactive} enhances the user's understanding of flooding depth, intensity, and impact with respect to its spatial surroundings. 
However, when analyzing a vast virtual scene at a high level-of-detail~(LoD), the limited field-of-view (FoV) of a single-screen desktop workstation can impel a user to engage in frequent pan-and-zoom interactions.  
This has motivated us to develop \textit{Submerse}, an application and system for flood visualization that harnesses the screen real-estate and visual acuity of large immersive visualization systems. 

Specifically for scientific analysis, large, high-resolution displays, such as CAVEs and immersive tiled displays have been shown to improve user performance in various visualization tasks~\cite{ball2007realizing, andrews2011information, laha2012effects}. 
In essence, utilizing large screen real-estate allows for a wider FoV.
This is particularly beneficial for applications such as flood visualization, where immediate access to local and global data context is vital for well-rounded analysis and decision-making.
Moreover, leveraging high-resolution graphics and high-performance capabilities of such facilities enhances the detail and granularity of the information.  

Submerse addresses 3 important design challenges.
First, visualizing simulation data coupled with elements of a geographic information system~(GIS) requires constructing and rendering large 3D meshes from data arranged on structured and unstructured grids.
Thus, a system infrastructure with low latency and real-time synchronization across display nodes is required for effective interaction. 
We have developed Submerse as a modular and scalable system that abstracts visualization design at the application level from its deployment and rendering across a multi-node multi-display system.
By providing the layout and cluster configuration of a display system, Submerse determines appropriate camera projections for each physical display in the setup, along with view-dependent data extents for rendering the virtual scene.
At the application level, Submerse renders the simulated flood as a 3D surface mesh with propagating waves to visualize flow direction. 
To optimize computation time and memory and to support LoD rendering, we represent the simulation data on an adaptive heightfield grid~\cite{cornel2015visualization,kim2009image} using a dynamic quadtree. 

Second, flood visualization applications require frequent data interaction.
Unlike a desktop setup, where applications have a graphical user interface and users have access to input devices, interaction in immersive settings usually adopts a natural mode, such as walking and hand gestures, and occasionally by using hand-held controllers. 
Submerse provides a unique focus-and-context interaction by integrating a hand-held video see-through augmented reality (AR) auxiliary (aux) display into the immersive setup. 
To achieve this, we have designed a physical-to-virtual spatial awareness protocol that augments additional information and visualizations atop the surrounding displays, facilitating local contextual analysis while simultaneously having access to global data. 
For an interactive user experience, Submerse includes a remote rendering protocol that offloads computation and rendering tasks from a commodity aux device to a visualization server.

Finally, effective navigation is an essential element for gaining a complete understanding of the scene.
This becomes additionally challenging for large and cluttered scenes, where multiple points-of-interest~(POIs) have to be observed.
Determining appropriate views and designing methods for navigation in immersive systems is an active ongoing area of research \cite{Bonaventura:2018:SVS}.  
To this end, we have developed an optimal camera view-finding algorithm that maximizes view coverage of user-defined POIs based on the layout of the display system.
Using the camera views, we additionally facilitate scene navigation by determining a camera path that 
smoothly transitions along subsequent viewpoints by connecting them using straight lines and quadratic B\'{e}zier curves such that it avoids collision with 3D scene objects, minimizes motion sickness for immersive environments~\cite{Hu:2019:RSS}, and maintains a camera look-at vector in the direction of the shoreline to ensure visibility of the coastal and inland flooding.

Submerse has been developed as a collaboration between computer scientists and atmospheric scientists and feedback from flood managers.
In this paper, we demonstrate the visualization of a flooding scenario in New York City~(NYC) and deploy it on the Stony Brook University Reality Deck (RD), the largest immersive gigapixel display facility in the world~\cite{papadopoulos2015reality}.
We evaluate our system based on feedback from workshops conducted with the NYC flood managers, meteorologists, domain scientists, graduate students in atmospheric sciences, and other stakeholders.

\section{Related Works}

\textbf{Visualization of Flooding Simulation Data:}
Tools for visualizing flooding scenarios generally feature a geographical map interface embedded with rasterized information from numerical simulation models~\cite{leskens2017interactive,noaa2,leedal2010visualization}.
However, it has been shown that 2D representation of flooding data limits a full spectrum understanding of multivariate geospatial attributes~\cite{vuckovic2021combining} and the severity of flooding scenarios, leading to ineffective preparation~\cite{mol2022after}.
To this end, Cornel et al.~\cite{cornel2015visualization} have designed a 3D application with visual encodings on virtual scene objects that provide qualitative and quantitative relationships between the objects and the flooding event.
In a follow-up work, the authors have developed a fast surface mesh reconstruction method using adaptive grids~\cite{cornel2019interactive}. This work also presents an algorithm for realistic water shading with depth perception cues.
Submerse extends the adaptive grid interpolation and wave synthesis techniques to dynamically reconstruct each grid submesh based on its view-dependent LoD and implements GPU-based tessellation. 
Our algorithms are designed to support the efficient rendering of large simulation data at interactive frame rates and to be scalable for distributed immersive visual display systems.

Particular to the VR domain, Oyshi et al.~\cite{oyshi2022floodvis} have developed FloodVis as an HMD-based 3D application that renders flood data along with scene elements, such as houses.   
Based on their evaluation and user study, the authors have identified system scalability and user experience in scene localization as limitations.
Our choice in designing Submerse as a system for immersive and tiled-display ecologies over HMD technology is primarily attributed to the ability to have a co-located collaborative space -- an HMD-based application will require additional multi-user approaches such as avatars, view synchronization, and effective user-to-user interaction methods; and data scalability -- current VR HMD technology is still limited in performing computationally expensive tasks along with high-quality rendering of large mesh.
Moreover, in contrast to existing flood visualization systems, Submerse provides a more wholesome framework for VR by presenting a novel interaction interface for collaborative spaces, and camera view-finding and path-planning algorithms specific to the domain requirements.  

\textbf{Interaction in Immersive Systems:}
Pose tracking has been the widely adopted technique to translate human gestures into interaction metaphors in immersive facilities \cite{davis2002lumipoint, ganser2006infractables, langner2018multiple}.
For direct manipulation of large data, touch-enabled surfaces have also been explored \cite{morris2006cooperative, malik2005interacting, wobbrock2009user}.
To utilize commodity hardware, Siddhpuria et al. \cite{siddhpuria2018pointing} have developed a system that integrates a personal smart device as a pointing device and Langner et al. \cite{langner2018towards} have explored interaction in large tiled-display systems using tracked mobile devices.
While these techniques apply interaction metaphors to the entirety of the displayed data, in Submerse, we take advantage of AR technology to provide a personalized-level exploration of the data.

The use of AR devices in immersive facilities has been previously explored.
Reipschlager et al. \cite{reipschlager2020personal} have used an AR head-mounted display (HMD) device to augment additional information on top of the display screens, and Nishimoto and Johnson \cite{nishimoto2019extending} have integrated AR HMD to extend a user's field of regard beyond the screen limits.
The former is designed particularly for infographics, and the latter for specific objects in the scene. 
In contrast, Submerse extends the ability to interact dynamically with the 3D virtual scene contents displayed on the surrounding screens.
For a seamless augmentation, we present a rendering protocol where the additional visualizations are geometrically corrected and integrated with the video see-through view of the virtual scene displayed on the immersive facility. 
While user-parallax view has been explored in video see-through AR \cite{barivcevic2012hand,hill2011virtual}, our approach does not adjust the entire video see-through content. 
Rather, unlike typical AR applications that render virtual objects onto the physical world, we render virtual objects that are meant to be augmented onto a virtual world that is displayed on a physical plane of display screens within a physical space.
In some sense, we have introduced an AR method for its referent in the VR space, displayed on a physical plane in the real world.
Moreover, to the best of our knowledge, our system is the first of its kind for individual-level exploration of flooding scenarios in immersive tiled-display visual systems.

\textbf{View-finding and Path-Planning:}
\label{sec:related:view}
Early on, Kamada and Kawai~\cite{Kamada:1988:SMC} have formulated rules for determining a good view of basic shapes by maximizing the projection of lines on the objects.
We refer the reader to a review by Bonaventura et al.~\cite{Bonaventura:2018:SVS} on viewpoint selection for polygonal data. 
More recently, information entropy ~\cite{feixas1999information} has been considered a measure for determining visible information.
Arbel et al.~\cite{Arbel:1999:VSN} have formulated view selection based on Shannon entropy maps by measuring object recognition in a monochrome view and  V\'{a}zquez et al.~\cite{vazquez2001viewpoint} modeled a maximization problem for object visibility.
Specific to landscapes, Stoev et al.~\cite{Stoev:2002:CSA} have addressed automatic camera positioning for terrains by maximizing scene depth and projected area.
In terms of navigation, Huang et al.~\cite{Huang:2016:TS} have introduced a camera motion design, particularly for urban scenes, and 
Lin et al.~\cite{lin2022capturing} have developed a framework for large-scale scene data perception and reconstruction. 
In contrast to existing works, our view-finding algorithm is specialized for flooding scenarios, addressing a two-fold optimization: 
(1) determining a view that maximizes information visibility of a POI while integrating surrounding features, such as terrain, occlusion due to scene objects, and its visibility from the coastline, and (2) adapting and improving the viewpoints based on a provided display configuration.
Using the determined viewpoints, we also develop a camera path control method for large-scale visualization, also taking into consideration the domain information.

\IEEEpubidadjcol

\section{Domain Requirements}
\label{sec:domain_reqs}

Submerse has been designed and developed as part of an effort by the Coastal Alliance Network for Visualization, Science, and Stakeholders (CANVASS).
The network aims to achieve a shared understanding of the problems and risks related to coastal and urban flooding among stakeholders, including scientists, emergency managers, and policymakers, by taking a creative approach in connecting predictive science and visualization.

To gather a set of system requirements, our team invited a focus group, comprising experts from the National Oceanic and Atmospheric Administration (NOAA), National Weather Services (NWS), NYC emergency managers, and social scientists, to understand the existing workflow of how simulations are analyzed.
Based on the NWS protocol, an informational orientation of an upcoming (simulated) storm was presented to the participants, using standard visualization tools \cite{noaa2}, that imitated a situation where experts engage in preparedness ahead of a weather event.
We observed their level of collaboration and engagement with the presented information, interviewed them regarding their decision workflow, and asked for feedback and suggestions regarding the visualizations effectiveness.  
Thereafter, we interviewed five atmospheric scientists to get insights into the facets of simulation data used in exploration, understanding, and analysis.
Following this process, we delineated the following requirements for Submerse:

\textbf{R1 -- 3D visualization of flooding scenarios:}
Analyzing numerical models that describe real-life environments around us makes atmospheric science a very visual field. 
While data collection methods and forecasting models have become better and more sophisticated, tools for their visualization are still primarily based on 2D projections. 
To better understand the flooding intensity, a visualization system is needed to stimulate a realistic perception of the flooding depth and severity with respect to its geographic surroundings and progression over time.  

\textbf{R2 -- A collaborative space for diverse stakeholders:}
A method to facilitate a fully integrative and collaborative space for stakeholders, such that users can experience a shared understanding as well as interact independently with the scientific findings of the climate catastrophe. 

\textbf{R3 -- Optimizing POI views and navigation in large scenes:}
Following an initial prototype, we noticed that one of the challenges of 3D visualization is scene navigation and adjusting the camera to achieve good views, especially in large urban scenes. 
To this end, for a set of input POIs, the system should assist with providing effective views for analyzing the progression of coastal flooding with minimal viewing occlusion.
For a collection of POIs, the system should aid in navigating the scene.

\begin{figure*}
    \centering
    \includegraphics[width = \linewidth]{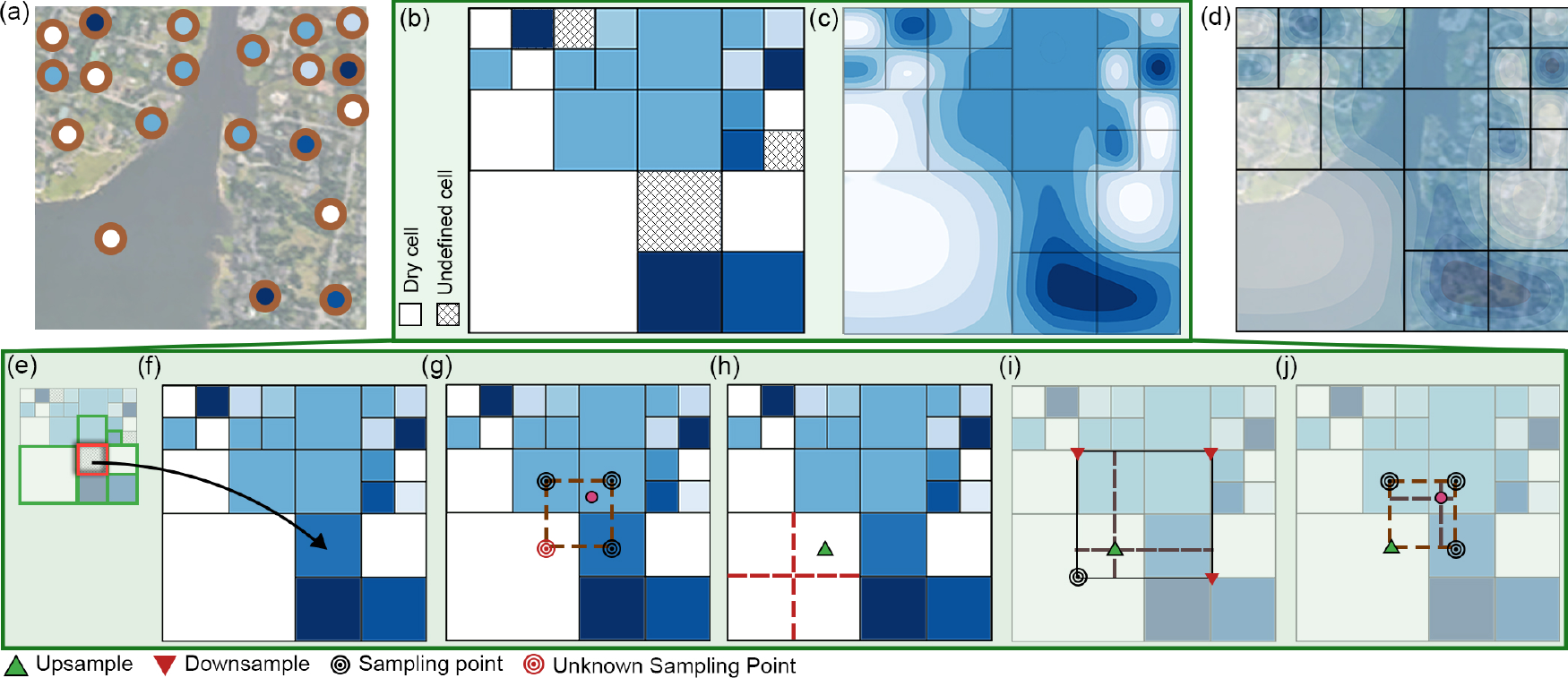}
    \caption{An example of constructing a water surface mesh using an adaptive grid. For flood datapoints defined on a geographic location (a), we discretize the data on a quadtree (b) and interpolate the values to construct a smooth and continuous field as shown in (c) and (d). A cell without a datapoint (e) is defined by averaging the values of its neighboring defined cells (f). To interpolate a value for the red point in (g), we apply bilinear interpolation on the cells that bound the point. However, since one of the sampling cells, marked in red, has a lower cell resolution, we divide the cell (h) and assign a value by upsampling. The upsampling is done by applying bilinear interpolation on the neighboring same-resolution cells (i). A lower resolution value is calculated by averaging the value of all its child higher resolution cells (downsampling).  Finally, a value for the point is constructed by applying bilinear interpolation on the same-resolution cells (j). }
    \label{fig:adaptive}
\end{figure*}

\section{Submerse}
Based on these requirements, we have designed Submerse as an application and system to visualize storm surge flooding simulations by specifically utilizing immersive display facilities.
For an input simulation data, we present an interactive visualization of the flooding scenario by integrating the flooding level into a to-scale 3D virtual model of the corresponding geographic area (\textbf{R1}). 
Coupling simulations with GIS and map-data elements allows a holistic understanding of the impacts of the hydrological variables with respect to contributing elements such as terrain, infrastructure, and population.
To this end, the virtual scene is additionally set up using GIS data, such as digital elevation map~(DEM) and satellite imagery, and supplementary scene objects, such as 3D buildings and reference 3D objects.
Sec.~\ref{sec:rendering} describes our method for processing the input simulation data to reconstruct an animated water surface, visualizing the flooding level and direction.

High-resolution display facilities such as CAVEs and immersive tiled-displays have been shown to enhance data realism and allow a more effective exploration of three- and higher-dimensional data by tapping into the human peripheral vision~\cite{laha2012effects} (\textbf{R1}).
Moreover, depending on the layout, these visualization systems present a natural space for multi-user collaboration and interaction (\textbf{R2}). 
In Sec.~\ref{sec:sys}, we present our Submerse system design for deploying the visualization application to a multi-node, multi-GPU display facility.
Specifically, based on the configuration and geometry of a display setup, Submerse manages synchronized instances of the application for each viewport in the system and optimizes computational and memory resources by determining view-dependent data extents.
This approach enables the abstraction of the visualization application, allowing it to be tailored based on scientific needs while supporting diverse visual display ecologies.

Beaudouin-Lafon~\cite{beaudouin2011lessons} has shown that in co-located sessions, researchers contribute better towards collaborative data consolidation and analysis when large displays are used for shared information while personal information mediums, such as laptops, tablets, and papers can be instantaneously accessed for focused information.
Gleaning from this, in Sec.~\ref{sec:aux_disp_app}, we describe our novel AR-based aux display system for immersive flood visualization, which we have designed to enable users to perform localized visual analytics seamlessly atop the displayed visualization (\textbf{R2}).

Finally, to facilitate an effective exploration of large scenes (\textbf{R3}), in Sec.~\ref{sec:cam_view}, we present an energy function that automatically generates optimal camera viewpoints for user-defined POIs, based on parameters such as the visualization display layout, POI visibility, and the distance from the user.
We generate a smooth camera path for navigating to the POI views using the set of calculated viewpoints.
Our camera path algorithm is designed to avoid occlusion with scene objects during navigation and aims to minimize VR sickness due to motion.

\subsection{Flood Simulation Rendering}
\label{sec:rendering}
The flood simulation data, which serves as input to our application, is a time series data of scalar water elevation values and 2D tidal velocity, discretized over a vast irregular grid of latitude and longitude positions.

\subsubsection{Quadtree-based Height-Field Reconstruction}
\label{sec:quadtree_recon}
Although using a high-resolution grid is desirable for generating a smooth and continuous surface, 
it has a high memory and performance overhead with increasing spatial extents.
To this end, we adopt adaptive height-field grid, which is a discretized representation of data on an unstructured grid of varying cell sizes~\cite{liang2011structured}. 
A surface mesh can be reconstructed for a particular resolution level by interpolating cell-centered values on the grid.
Fig.~\ref{fig:adaptive} illustrates how this process is applied to our application.
For a collection of water elevation datapoints (Fig.~\ref{fig:adaptive}(a)), we discretize the data using a quadtree datastructure (Fig.~\ref{fig:adaptive}(b)).
The main advantage of using a quadtree is its low memory footprint, which is particularly important for limited GPU memory, and its hierarchical data representation, enabling LoD rendering.
Each leaf node in the quadtree corresponds to a cell of the height-field grid and the value of each cell is defined using the datapoint value bounded by the leaf node (illustrated using shades of blue in Fig.~\ref{fig:adaptive}(b)).
To enable LoD rendering, we assign parent nodes to have a value equal to the average value of their children.   
Finally, using a regular triangular mesh, we render a  water surface where the height of each vertex is determined by applying 2D bilinear interpolation on the four cells that enclose the vertex (Fig.~\ref{fig:adaptive}(c)).

Notice in Fig.~\ref{fig:adaptive}(b) that in sub-dividing a quadtree node, a child node without a datapoint can result in an \textit{undefined} height-field cell value.
Moreover, the simulation output can also have \textit{dry} datapoints, that is, the spatial location does not flood for certain timepoints, and in some cases, the entire duration of the simulated flood (this is specifically to define a boundary for inland spatial locations beyond which no flooding occurs).
For dry cells, we use the average height of the underlying terrain obtained from the input DEM, whereas for the undefined cells, we assign an average value of all the defined cells in the neighborhood.

To correctly interpolate height values from varying cell sizes, we modify the adaptive reconstruction introduced by Cornel et al.~\cite{cornel2015visualization}.
In their method, if a neighboring cell is of a different size, an imaginary cell with an equivalent size is reconstructed by upsampling (for a higher resolution level) or downsampling (for a lower resolution level).
In our variation, the downsampled value is simply the averaged value assigned to the corresponding node parent, as in our quadtree reconstruction above.
To reconstruct an upsampled value, a similar 2D bilinear interpolation is applied using the lower resolution values of the four bounding cells.
Figs.~\ref{fig:adaptive}(e)-(f) show this process schematic representation.
One caveat to this approach is that the cell neighbors can differ by, at most, one level of resolution.
We implement this constraint in our quadtree construction step.

We furthermore extend this technique to facilitate LoD rendering.
Although the hierarchical datastructure of quadtrees supports LoD by design, we restrict the lowest renderable resolution of a sub-tree to be the node whose children belong to the same type, that is, all four children should either be flooded or dry.
This condition prevents dry cells from being incorrectly estimated to have flooding.
Note that for leaf nodes that correspond to undefined cells, their extrapolated flooding value is used to propagate to the parent node, thus categorizing it as a flooded node.

\begin{figure}
    \centering
    \includegraphics[width = \linewidth]{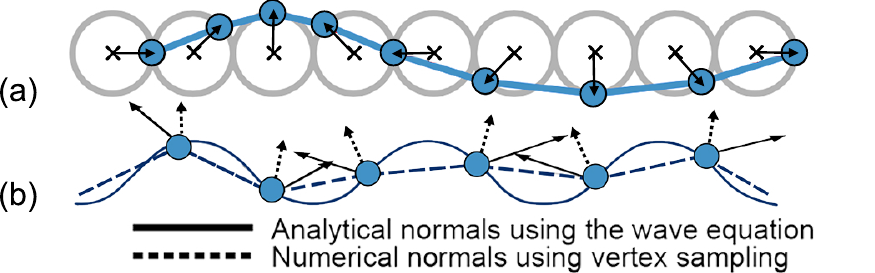}
    \caption{(a) shows vertex motion following a Gerstner wave, and (b) illustrates the difference in calculating analytical and mesh normals.  }
    \label{fig:Gerstner_wave}
\end{figure}

\subsubsection{Flooding Direction Visualization}
To visualize tasks that have a behavioral attribute of flow, for instance, inundation, information cannot be perceived by observing a static mesh.  
This becomes particularly true when time-series data is visualized while paused for a single instance. 
A natural instinct then is to repetitively toggle back and forth to observe change and variation.  
Moreover, visual metaphors such as glyphs may add to the complexity of the visualization~\cite{cornel2015visualization}, especially amongst diverse stakeholders.

To enable an intuitive perception of flow for a single timepoint, we animate the water to emulate wind-driven surface waves.
Typically in computer graphics, realistic waves are synthesized using the Navier-Stokes equation.
However, for a more efficient and simplified approach, we apply Gerstner wave synthesis model~\cite{hinsinger2002interactive}  using velocity vectors from the simulation data. 
Gerstner waves produce nearly realistic-looking waves based on a fundamental observation that, unlike a particle system where particles move with the wave, mesh vertices can be displaced along a trochoidal function.
Specifically, as the crest of a wave approaches, in addition to being vertical displacement defined by a sine function, a point moves horizontally towards the direction of the crest to simulate the phenomenon of water particles filling up the space under the crest.
Conversely, as the crest passes, the point slides back to its resting position on the horizontal axis. 
A schematic representation of this displacement is shown in Fig.~\ref{fig:Gerstner_wave}.

For a vertex point $P_{xyz} = (x,y,z)$ on a surface mesh, its displacement along a Gerstner wave is calculated as follows:

\begin{equation}
\label{eq:singlewave}
W(P_{xyz}, \Vec{v}, \lambda, t) = 
\begin{cases}
P_{xy} + \frac{s}{k} \cos\left(kD - ct\right) & \textrm{for } (x,y)\\
P_z + \frac{s}{k} \sin(kD - ct) & \textrm{for } z
\end{cases}
\end{equation}
\begin{equation}
    k = \frac{2\pi}{\lambda} \textrm{,} \quad  D = P_{xy} \cdot \Vec{v} \textrm{,} \quad c = \sqrt{\frac{g}{k}}
\end{equation}
where $\lambda$ is the wavelength, $g$ is gravity, $\Vec{v}$ is the 2D unit vector of the simulation tidal velocity, $s$ is a defined wave steepness parameter, and $t$ is the animation time interval.

Directly sampling wave orientations from the corresponding quadnodes results in tiling artifacts, and rendering a single uniform train of waves traveling across the water surface presents a non-realistic feature. 
Existing works have suggested superimposing multiple waves sampled from increasing spatial extents, especially for open water bodies~\cite{thon2000ocean, hinsinger2002interactive}.      
This not only maintains a degree of realism but also allows the visualization of a regional principal flow as well as local flow direction.

In our implementation, we use the overlapping bounds suggested by Cornel et al.~\cite{cornel2015visualization}.
An imaginary tile, $T$ around $P_{xy}$, for a spatial extent spanning $\lambda$, is defined  as $T = [x_0, x_1] \times [y_0, y_1]$, where
\begin{equation}
    (x_0, y_0) = 2 \lambda \left\lfloor\dfrac{(x,y)}{2\lambda}\right\rfloor \textrm{,} \quad (x_1, y_1) = (x_0, y_0) + (2\lambda, 2\lambda)
\end{equation}
A wave for $T$ is then calculated using its center position and an average tidal direction sampled from each of its four corners. 
To get an overlapping region, $T$ is shifted horizontally and vertically by $\lambda$ such that the centers of the four tiles $[T, T + ( \pm\lambda, 0), T + (0, \pm\lambda), T + (\pm\lambda, \pm\lambda)]$ enclose $P_{xy}$ in a box of length $\lambda$.
Thus, a wave $W_\lambda$ is calculated using the following bilinear interpolation at each tile center:
\begin{equation}
    \alpha = \frac{1}{2} - \frac{1}{2}\cos\left(\frac{\Delta x}{2\lambda}\right) \textrm{,} \quad \beta = \frac{1}{2} - \frac{1}{2}\cos\left(\frac{\Delta y}{2\lambda}\right)
\end{equation}
where $\Delta x$ and $\Delta y$ are the distance of $P_{xy}$ along the $x$ and $y$ directions with respect to the tile centers. 
For superimposing multiple waves of increasing spatial extents, waves for $n$ different wavelengths are generated using $\lambda_i = r_i2^i\lambda_0$, where
$\lambda_0$ is the shortest considered wavelength and $r_i \in [0.8, 1.0]$ is a random value to avoid frequency doubling. 
The final vertex displacement is determined using:
\begin{equation}
\label{eq:multiple_waves}
    W = W_{\lambda_{0}} + \sum_{i=1}^{n}W_{\lambda_i}
\end{equation}

Since the waves are generated on a triangular mesh with discrete intervals, we use analytical normals instead of vertex normals for continuous lighting and shadow effects (see Fig.~\ref{fig:Gerstner_wave} for this comparison).
The normal $\Vec{n}$ for each vertex is the spatial derivative of Eq.~\ref{eq:singlewave}:
\begin{equation}
    \Vec{n} = 
    \begin{bmatrix}
    1 - D^2_x s \sin f \\
    D_x s \cos f \\
    - D_x D_z s \sin f
    \end{bmatrix}
    \times
    \begin{bmatrix}
    - D_x D_z s \sin f \\
    D_z s \cos f \\
    1 - D^2_z s \sin f
    \end{bmatrix}
\end{equation}

Finally, we address two implementation details.
First, an averaged neighborhood tidal velocity is assigned to an \textit{undefined} node, however since \textit{dry} nodes do not have tidal velocities, we set $W_{\lambda_{0}} = 0$.
Second, to reduce loss of tidal direction due to averaging, we impose a further constraint for determining the lowest resolution LoD subtree, such that the angles between all pair-wise unit vectors of the subtree children should not exceed 15\textdegree.

\subsection{Immersive Display System}
\label{sec:sys}
Submerse is intended to be generalized for any display layout and configuration -- a single desktop to a large and immersive setup, a single node to a multi-node, multi-GPU configuration. 
For an input display setup, we (1) determine the scene data extents for each display and use dynamic quadtrees to allow interactive data updates, 
(2) set up a localization and rendering protocol for the auxiliary display, and (3) configure optimal camera views for marked POIs.

A display system is defined using the position, orientation, and dimensions of each screen in the layout, along with a \textit{head} position.
The \textit{head} is primarily a reference point that represents a unified position of the display system in the virtual scene and is used to calculate the camera projection matrix for each display.
At a systems level, Submerse is designed as a client-server architecture, where the server manages scene synchronization and sends interaction commands across the client (display) nodes.

\begin{figure}
    \centering
    \includegraphics[width = \linewidth]{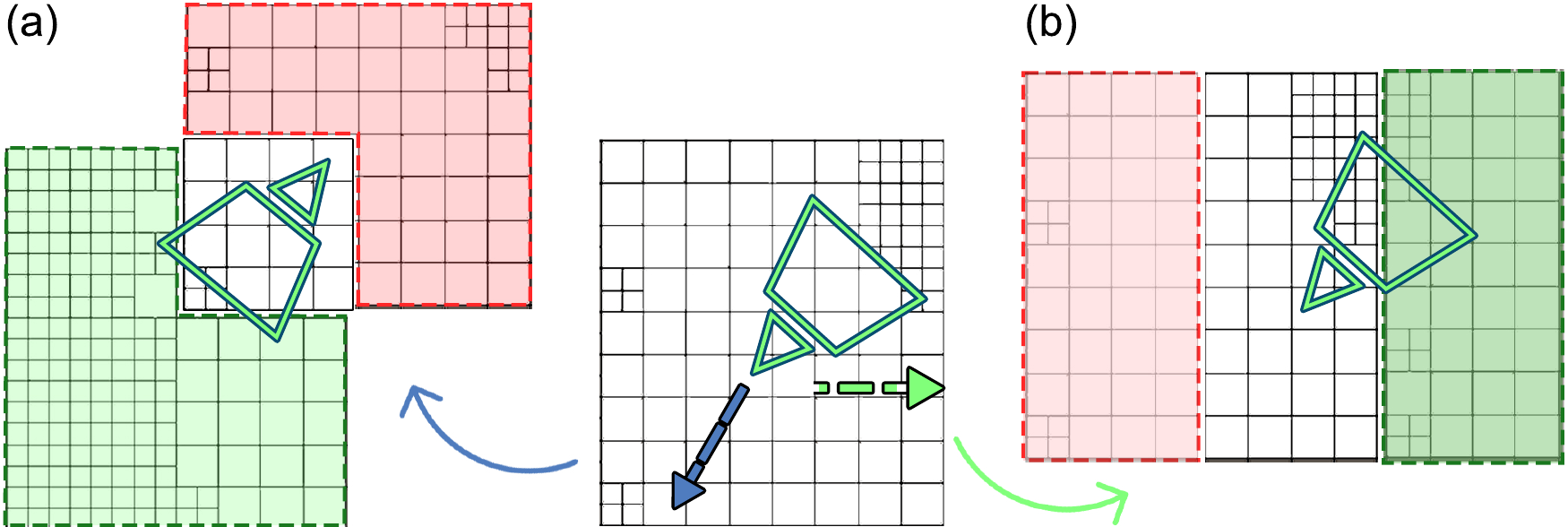}
    \caption{Two examples (a) and (b) of our dynamic quadtree update based on camera motion. Red represents the deleted quadrants and the newly added nodes are shaded in green. }
    \label{fig:dynamic_quad}
\end{figure}

\label{sec:data_extents_and_update}
To optimize data retrieval, we have designed our quadtree to contain data for spatial bounds proportional to the display's viewing frustum, which is dynamically updated during scene navigation. 
Specifically, the bounds of a quadtree are initialized such that the virtual position of the display lies in the center of the quadtree and it spans the maximum horizontal extents of the camera-view frustum pyramid $+$ a buffer region.
The camera view frustum for a screen is determined using its dimensions, relative position from the \textit{head}, and camera parameters.

For maintaining an interactive framerate during scene exploration, we have designed the first-level quadrants of the quadtree (children of the root node) to be dynamic.
That is, the quadrants are selectively updated based on the camera movement direction. 
By tracking the displacement of the physical screen in the virtual scene, we determine which quadrant needs to be updated as the screen approaches close the quadtree outer boundary.
Consequently, instead of constructing an entire quadtree with new spatial extents, we discard the quadrants that are beyond a threshold distance from the screen position, rearrange the pointers of the preserved quadrants to reflect the new root bounds, and populate nodes for the new quadrants.
In Fig.\ref{fig:dynamic_quad}, we illustrate this process using two examples of camera movement.
In our implementation, the quadnode updates are performed in parallel using multi-threading and are either triggered as the user manually navigates through the virtual scene or scheduled sequentially based on a predefined camera path generated using our algorithm described in Sec.~\ref{sec:cam_view}.

\begin{figure*}
    \centering
    \includegraphics[width = \linewidth]{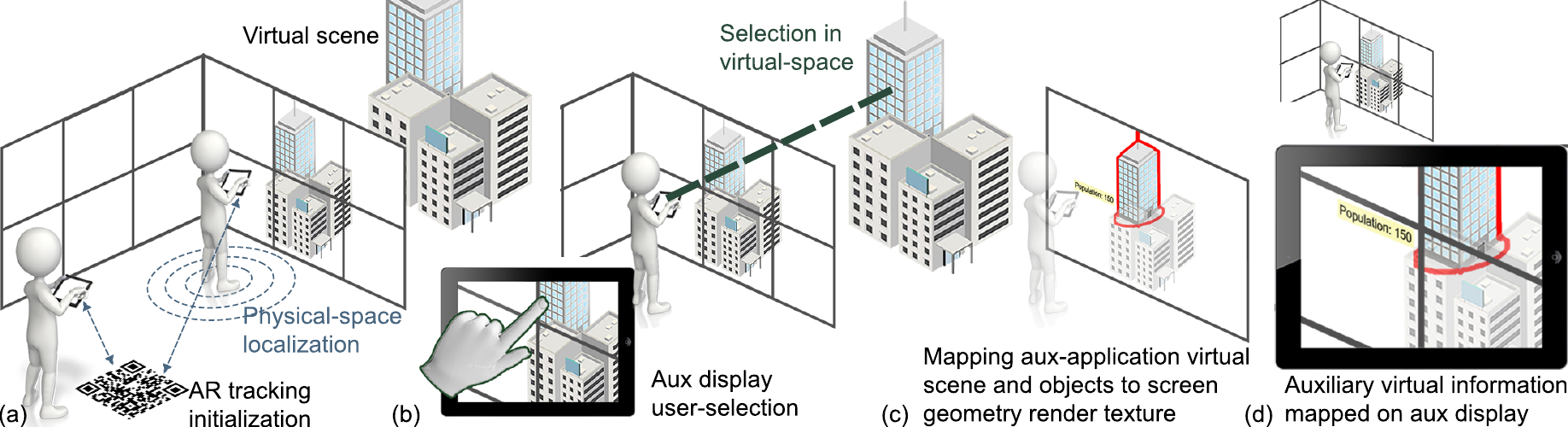}
    \caption{System protocol of our AR-based auxiliary (aux) display. (a) Localization is initialized by registering an image marker, placed at the physical location of the defined \textit{head}. (b) The tracked position and orientation of the device are transformed from the physical to the virtual space, and data interaction is supported by casting a ray from the physical device position to the virtual scene. Selection and depth are performed by tapping and swiping gestures. (c) Complementary visualizations are rendered on a temporary render texture to accurately augment additional virtual objects on the aux display, as shown in (d).}
    \label{fig:aux_disp_sys_apps}
\end{figure*}
\subsection{Auxiliary (aux) Display}
\label{sec:aux_disp_app}

For analysis and decision-making of large datasets, visualization applications typically provide users with tools that can systematically present a global overview of the data as well as a detailed depiction of a selected local area.
To this end, we have designed a novel mixed-reality focus+context technique where the global data is projected on the surrounding display facility while localized contextual visualizations are shown on a hand-held AR-enabled device (which we term an auxiliary display).

\subsubsection{Device Setup}
One of the essential features of our 3D visualization is providing terrain and scene context to the flooding simulation, creating a virtual scene for the scenario.
Synonymous with using AR in physical reality, we introduce a method for using an aux display to augment virtual content onto dynamic VR scenes.
Since visual display facilities comprise a physical space that facilitates user movement and display screens that act as a ``window'' looking into the virtual scene, the aux system must accommodate tracking and localization in both the physical and virtual worlds.
This includes tracking the device position and orientation in the physical space and aligning it with the virtual camera that is used to render the virtual scene.

To establish a common reference point, an initialization tracker is placed at the planar position (floor) of the defined \textit{head}.
In our implementation, we permanently place an image-based marker onto the position in the display facility.
This allows the AR system to recalibrate its tracking whenever the image marker is in view and enables multiple devices to connect to the system.
By leveraging recent advances in AR libraries, our Submerse aux device application performs on-device physical space localization with respect to the marker.
Subsequently, by connecting the aux device to the configured server node, the Submerse server application uses the input display facility layout and the device position and orientation to resolve the physical-to-virtual transformation necessary for determining interactivity and rendering the complementary visualizations on the aux display.
Fig.~\ref{fig:aux_disp_sys_apps} presents a schematic representation of the system protocol. 

\subsubsection{Integrating AR for VR in Physical Space}
As the user moves in the display facility, the device position and orientation are persistently communicated to the server. 
For data interaction, a raycast from the center of the tracked device is used to resolve object intersection in the virtual space.
Swiping and tapping gestures on the device are used to determine the ray depth and selection, respectively.

Here we address an important challenge related to AR-based rendering of the visualizations on the aux display. 
The physical-to-virtual mapping introduces scale disparities since two cameras are involved, the aux display camera for providing the see-through feed and the Submerse virtual camera.
To resolve this, we utilize render textures on the server.
First, based on the device localization, the designed complementary visualizations are rendered in an empty virtual scene with extents similar to the flooding scene.
Next, we use the Submerse camera parameters for baking a view of the rendered visualization onto a virtual texture that replicates the size and layout of the display facility.
Thus, instead of re-rendering the virtual scene that is displayed globally on the surrounding screens and visible through the video see-through feed, the aux display rasterizes the required visualizations from the render texture based on the device camera parameters. 

Finally, given a mobile device limited GPU and compute resources, user experience can be significantly affected when rendering complex scenes and managing large datasets. We have implemented a web-based remote rendering module into our system to offload this burden from the aux device.
As a result, the rendering is performed on the server, and the final result is a series of textures received by the device over the network.

\subsubsection{Submerse Auxiliary (aux) Visualizations}

\begin{figure}
    \centering
    \includegraphics[width = \linewidth]{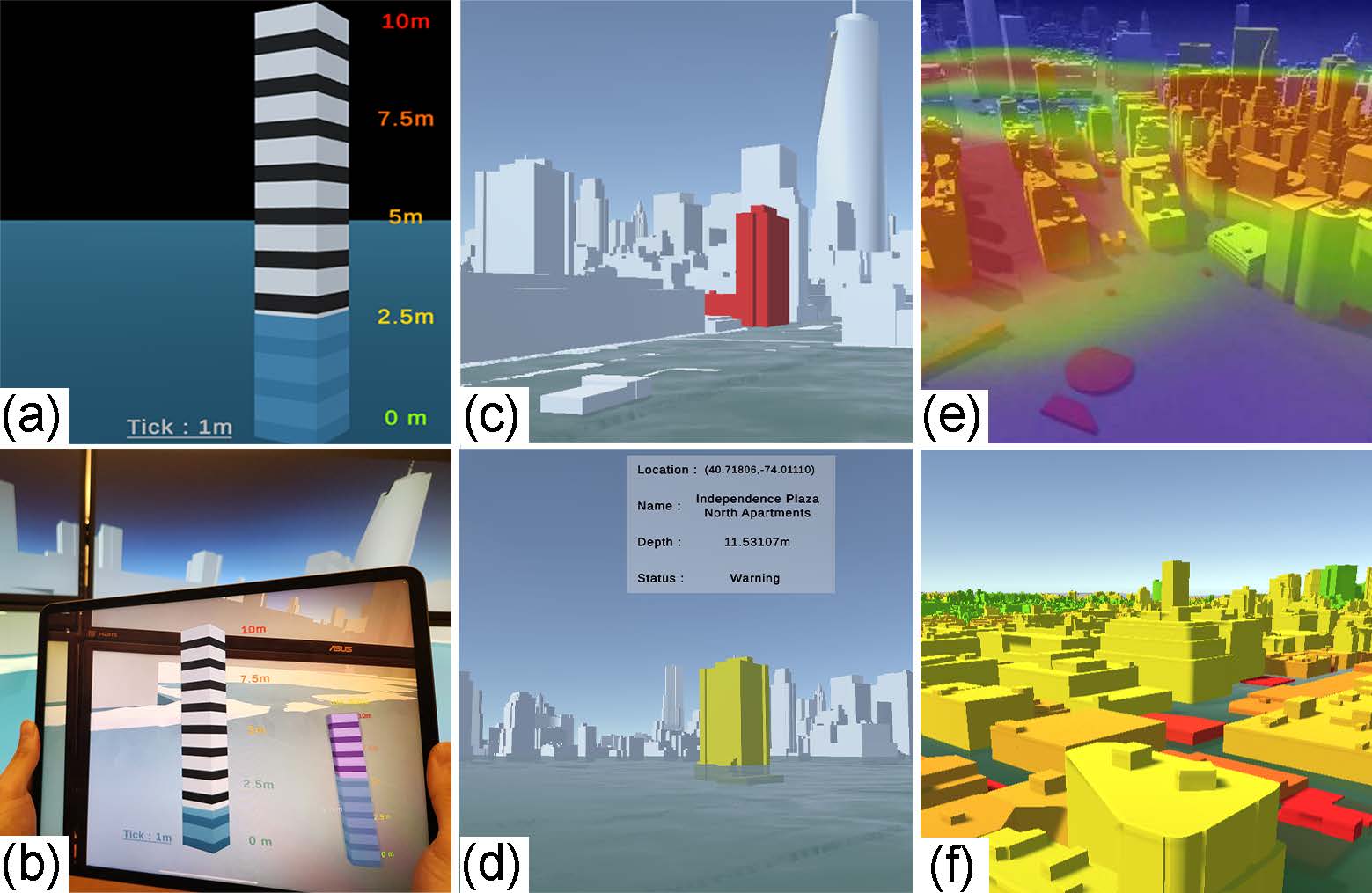}
    \caption{Examples of Submerse auxiliary (aux) display visualizations.}
    \label{fig:aux_display_apps}
\end{figure}

Based on domain expert input during our system requirements gathering process (in Sec.~\ref{sec:domain_reqs}), we have implemented the following visualization and interactions for focus+context on the aux display:

\textit{Dip-stick metaphor:}
Meteorologists often prefer to note quantitative measurement values. 
Desktop-based applications achieve this by providing a legend or designing a mouse-based tool-tip to display the depth value.
To adopt a natural way of measuring depth, we render a ruled dip-stick on the aux display, as shown in Figs.~\ref{fig:aux_display_apps}(a) and (b), that show water markings on the measurements when the device is pointed to the water surface mesh.

\textit{Selection and occlusion removal for buildings:}
Since the aux display can complement interaction metaphors such as selection, we provide users the ability to point and select specific buildings to view additional information.
Fig.~\ref{fig:aux_display_apps}(c) demonstrates such an example where for a selected building in red, we change its color to represent the flooding level severity relative to its height (shown in see Fig.~\ref{fig:aux_disp_sys_apps}(d)), and a label with information such as its name, location (address, along with latitude and longitude), and any additional related information provided by the user.
Moreover, to enhance visibility, we remove scene objects and buildings that occlude the selected building in the aux display camera view and in a radius around the building. 

\textit{Visualizing additional contextual information:}
For any domain-specific additional data, we allow users to (1) select the 3D scene object class for which the data is to be visualized, namely, water surface mesh, 3D buildings, or a new plane spanning the scene extents, and (2) configure a transfer function for the rendering. 
Figs.~\ref{fig:aux_display_apps}(e) and (f) are examples of this feature, demonstrating a heatmap of NYC population density atop the flooding scenario and a transfer function of flooding level severity relative to the building number of floors, respectively.

One of the advantages of Submerse as a modular system is that developers can design and implement custom visualizations and interaction metaphors based on their collaborating domain expert requirements. The system will accordingly deploy it to aux devices and integrate it with the immersive display layout.

Moreover, users can share their visualizations in a multi-user setting to facilitate collaborative discussions. 
This is shown in Fig.~\ref{fig:aux_display_apps}(b), where the dip-stick highlighted in pink is shared by another user. 
When a user selects to share a visualization, the selected object and its corresponding visualization type are sent to the server node, which then renders it to the render texture of the remaining connected users. 
In its current state, we support one-to-all sharing. 
For future work, we plan to present a study and design of a scalable and memory-efficient collaboration protocol for  AR aux display devices in immersive ecologies. 

\subsection{Camera View Finding and Navigation Trajectory}
\label{sec:cam_view}
Now we describe the objective function that we have designed to automatically determine camera viewpoints for user-input POIs, based on factors that maximize visibility with respect to the virtual scene and physical display layout, followed by our algorithm to generate a path for navigating through the viewpoints.
A POI is defined using its 3D position and a scalar bounding radius, and a camera viewpoint, $v$ is defined using its 3D position, and yaw and pitch angles.
For all the terms described below, $E_X$ is a large-valued penalty score to constrain the system to satisfy the respective user-defined threshold value or condition.

\textbf{View coverage:}
To anchor a POI in the center of the camera view frustum, we define view coverage, $R(v)$, as the combination of the eccentricity of a POI's projection on the camera view and the alignment between the camera \textit{up} vector and $v$:
\begin{equation}
R(v) = 
\begin{cases}
    E_R, & \text{if $a_p > A_p $} \\
    \omega_u \left\Vert\frac{2 a_p}{\pi}\right\Vert^2 - \left(\overrightarrow{pq} \cdot \vec{d} \right), & \text{otherwise}
    \end{cases}
\end{equation}
where $\omega_u$ is a weight to control the relative influence between eccentricity and viewing direction alignment, $a_p$ is the pitch angle of the camera pose, $p$ and $q$ are the positions of the camera and POI respectively, $\vec{d}$ is the camera look-at direction, and $A_p \in [0, \pi/2]$ is a user-defined maximum allowable pitch angle for the camera.

\textbf{POI visibility:}
To ensure effective visibility of a POI, a candidate view is penalized when the POI is occluded:
\begin{equation}
O(v) = 
    \begin{cases}
    E_O, & \text{if this POI is occluded} \\
    0, & \text{otherwise}
    \end{cases}
\end{equation}
We use a discretized visibility grid to check if the POI is occluded by any cell containing a 3D scene object or another POI when projected onto the camera view.
A visibility grid cell is defined using a cuboid with a height equal to the maximum height of the 3D objects occupying the cell bounds (except the water mesh).
Additionally, we define a term $D(v)$ to control the closeness of the POI to the display:
\begin{equation}
    D(v) = 
    \begin{cases}
    E_D, & \text{if $\frac{ ||q - p||^{2} }{k} > K$ }\\
    \frac{ ||q - p||^{2} }{k}, & \text{otherwise}
    \end{cases}
\end{equation}
where $k$ is a constant to normalize the distance calculate and $K$ is the user-defined maximum allowable distance of the POI from the camera.
A larger $K$ achieves a resultant view covering multiple POIs in a single camera view, whereas a smaller value results in a \textit{street-view} (in conjunction with a small coastal vicinity weight described in Eq.\ref{eq:optimization}).

\textbf{Coastal vicinity:}
Often, for POIs on the coastline, it is preferred that the progression of the coastal flood approaching the POI may be observed. 
We model this as location preference, $L(v)$:
\begin{equation}
    L(v) = 
    \begin{cases}
    0, & \text{if in offshore locations on map} \\
    E_L, & \text{otherwise}
    \end{cases}
\end{equation}

\textbf{Final objective:}
To find a set of optimal viewpoints, $C$, for $M$ POIs, we define the following objective function with customizable weights: 
\begin{align}
    \begin{split}
    C &= \argmin E(v) \\
    E(v) &= \sum^{M} {(\Omega + R_{i}(v) + \omega_o O_{i}(v) + \omega_d D_{i}(v) + \omega_l L_{i}(v) )}
    \end{split}
    \label{eq:optimization}
\end{align}
We use $\omega_u=0.5, \Omega=10.0, \omega_o=1.0, \omega_d=1.0, \omega_l=0.50$ and the penalties $E_X = 100.0$ to generate visualizations shown in our results (Sec.~\ref{sec:results}).
However, the weights in the viewpoint optimization can be adjusted to reflect the importance of each optimization term.

For $N$ displays, the total objective is calculated as:   
\begin{align}
    \begin{split}
    E(v) =  \sum^{N}_{i} {\sum^{N^{'}}_{j}{E_i(v_j)}} \quad \quad N^{'} = \{x \in N \mid x \neq i\}
    \end{split}
    \label{eq:ndisp_objective_function}
\end{align}
Eq.~\ref{eq:ndisp_objective_function} accumulates a score for all displays in the layout by rotating a candidate camera viewpoint $v$, along its look-up axis, with the relative yaw angle between display pairs.
This allows us to obtain a set of minimum camera viewpoints for multiple POIs for the display layout.
We demonstrate this using 2 layouts in our results (Sec.~\ref{sec:results}).

The terrain elevation and building heights create a complex search space for resolving minima since the derivative of a sample point contributes little to locating an optimal candidate.
Therefore, we chose Particle Swarm Optimization (PSO) \cite{Kennedy:1995:PSO}, a derivative-free optimization method that does not depend on a sampled gradient or covariance matrix of the sample set. 

Finally, given a set of viewpoints, we calculate a trajectory between viewpoint pairs. 
When possible, a straight line segment between viewpoint pairs is generated to reduce unnecessary rotation that can lead to VR sickness~\cite{Hu:2019:RSS}.
However, to avoid occlusion during navigation, where a straight-line trajectory is interrupted, we transform the path to a quadratic B\'{e}zier curve.
We allocate the curve's middle control point using a vector rooted at a point off the coast or a further point that results in no intersection.
Choosing an off-the-coast point is based on a preference expressed by domain experts to visualize coastal flooding while navigating, a disaster analysis protocol commonly undertaken by emergency managers. 
To reduce VR sickness due to sharp curvature, we set the vector length to be three times the distance between the center point and the straight line.

\begin{figure*}
    \centering
    \includegraphics[width = \linewidth]{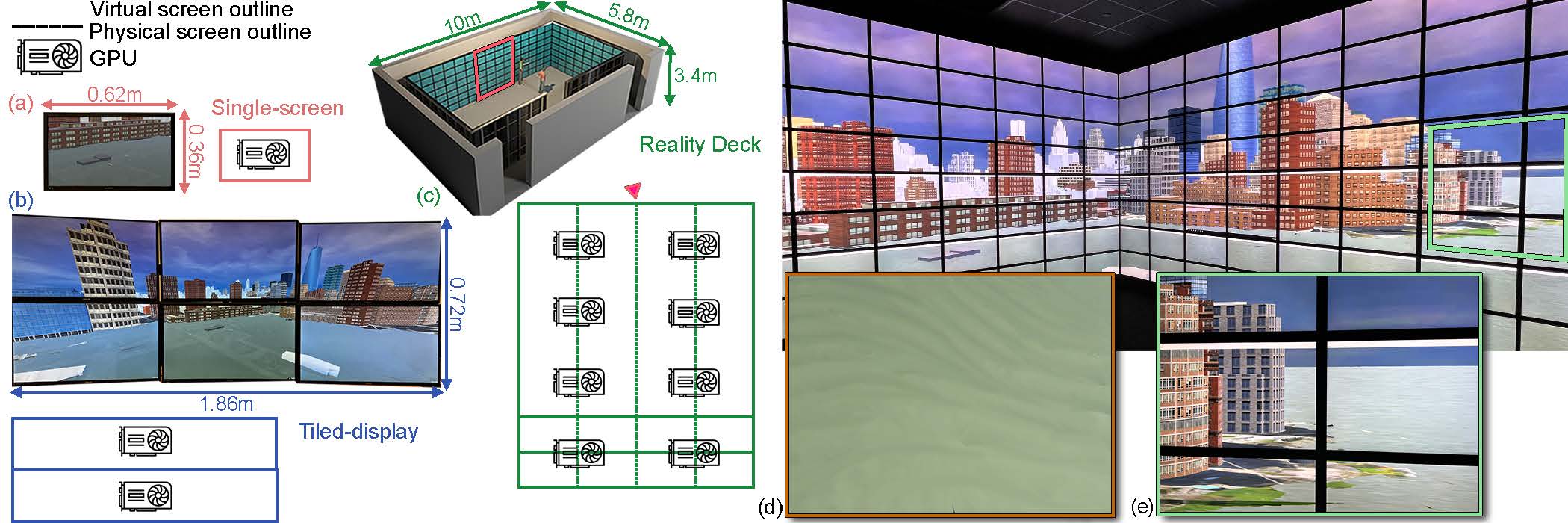}
    \caption{We demonstrate Submerse on three display systems: (a) single-screen, (b) tiled-display, and (c) Reality Deck, a room-sized gigapixel immersive facility. Inset (d) shows a result of our wave synthesis, and (e) illustrates the detail that can be resolved by approaching the facility walls. }
    \label{fig:disp_results}
\end{figure*}
\section{Implementation Details}
\label{sec:implementation}

Submerse is developed using Unity 3D~\cite{Unity3D}.
The modular and drag-and-drop interface provides an interactive framework for configuring, amending, and adding components based on application needs.
We use Unity AR Foundation to perform the aux device localization and WebRTC~\cite{webrtc} for implementing remote rendering.

To set up the display system layout, we interactively define the physical width, height, and relative placement and orientation of the virtual screens using quads as a visual guide (see Fig.~\ref{fig:disp_results}). 
A virtual screen is a canvas that can comprise either a single or multiple monitors, typically configured at the operating system or the GPU driver level.
For each virtual screen, a set of attributes needs to be configured, namely, for multi-node systems, its network address, and for multi-GPU systems, the GPU port number.
Using these configurations, we can launch instances of Submerse on the respective display nodes for the respective virtual screens.
A server application then connects to each client instance and performs its synchronization and interaction tasks in parallel, implemented using Unity's networking protocol.

To ensure an interactive framerate, we compute the vertex height interpolation and wave generation on the GPU. 
At the application startup and during the scene update, a triangular mesh with intervals equal to the quadnode length is generated.
By initializing a coarse mesh on the CPU, we optimize the number of triangle draw calls on the GPU by performing hardware-level tessellation.
To this end, we apply two tessellation passes on each triangle.
For the first pass, we subdivide a triangle based on its distance from the camera. 
The tessellation factor is maximum when the camera is at a minimum defined distance from the water surface (e.g., our default value is 100 meters), and gradually decreases until it reaches a maximum defined distance (e.g., our default value is 2000 meters).
If the triangles in the initial mesh have largely varying edge lengths, the output from the distance-based tessellation pass will result in smaller triangles being tessellated more than larger triangles. 
Thus, in the second tessellation pass, the triangles in the camera frustum are refined to have a maximum allowable edge length.
Although we empirically set this value to be one-tenth of the smallest quadnode length, this can be changed based on available GPU resources. 
Finally, to avoid artifacts at the boundaries of the quadnodes due to tessellation, we generate a fine triangular mesh at the boundaries and limit tessellation to the internal mesh of a quadnode.

A limitation of using height-field grids is that information is cell-centered rather than spatially localized.
For new vertices generated during tessellation, it becomes challenging to check for intersections when the wave is only partially visible on the terrain. 
Thus, in our shader design, we cache the interpolated height of each vertex, along with its crest and trough heights, and perform a depth test using a depth buffer.
Since the wave amplitude in our design is solely to induce a perception of flow and does not hold numerical significance, we perform the following shader operations: (1) if the cached interpolated height is projected higher in the depth buffer than other scene objects and either the crest or trough is occluded, the amplitude is reduced to a minimal value (in our case, 1mm), (2) if the interpolated height is occluded in the depth buffer, it is shaded transparent.

Our optimal camera viewpoint search is in a 5D  space (3D position and pitch-yaw orientation). Each particle in the PSO represents one display pose candidate. Since our optimization problem is not under-constrained, it is guaranteed to converge to a single solution. 
Defining the number of particles is a trade-off between the computation cost of dispatching more particles in each iteration and the number of iterations. 
We set a relatively small particle number of 1000  because our objective function yields light-weighted computation cost within each particle.
Thus, allowing us to increase the number of iterations at a lower time cost.

\section{Results and Evaluation}
\label{sec:results}

We demonstrate Submerse for a simulated flooding scenario of superstorm Sandy impacting NYC~\cite{colle2015exploring}, visualized on three kinds of display modalities: a single-screen desktop, a single-wall multi-GPU tiled-display setup (tiled-display), and the Stony Brook Reality Deck (RD), a room-sized 1.5 gigapixel-resolution immersive display facility. 
The specifications and layout for each display system are illustrated in Figs.~\ref{fig:disp_results}(a)-(c), respectively.
All three systems are equipped with NVIDIA Quadro RTX6000 GPU hardware and Intel Xenon 2.80GHz CPU with 128GB of RAM.

The flooding data is generated using the ADCIRC simulation model and spans the NYC area ($302.6 \text{mi}^2 $), containing 908,000 datapoints $\times$ 288 timepoints (30 mins interval).
To construct a NYC virtual scene, we used publicly available datasets~\cite{nyc_opendata} to integrate a DEM, which we texture map using the highest available resolution aerial NYC images~\cite{osm}, and 3D buildings. 
For this scenario, Submerse managed a total data size of 15GB (excluding aerial textures, which we directly streamed from the internet at runtime). 

In terms of performance, the startup time for the single-display, tiled-display, and RD, at the highest LoD, were 66s, 110s, and 145s, respectively. 
This accounts for 3D object loading, quadtree construction, and floodwater surface rendering. 
The texture mapping was performed in parallel, streamed directly from the internet, and thus not included in the time performance.
Once initialized, Submerse synthesized and rendered animated waves at 40 frames-per-second (fps).
Fig.~\ref{fig:disp_results}(d) shows an example of waves synthesized using our approach.
During manual scene navigation, updating the quadree using our dynamic approach maintained 25-30 fps.
However, when using our automatic camera path algorithm the path is pre-calculated, the quadtree is updated preemptively in parallel, and thus the fps is not affected.

\begin{figure}
    \centering
    \includegraphics[width = \linewidth]{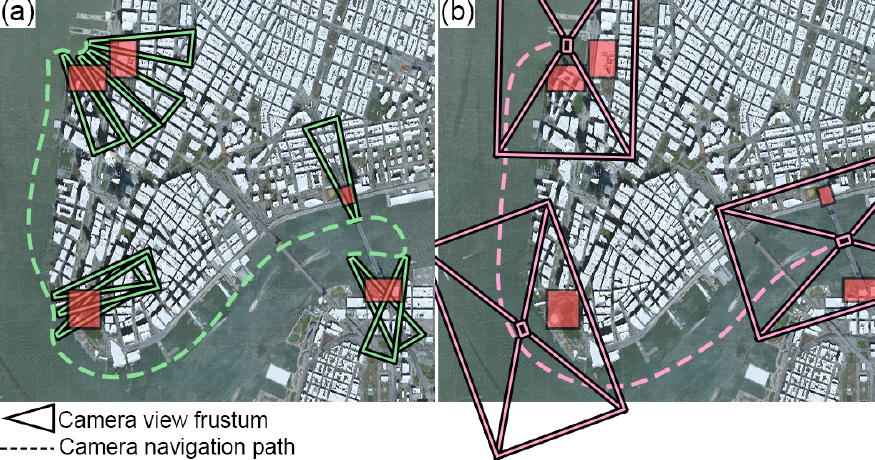}
    \caption{Camera views and navigation path generated for (a) a single screen and (b) four-walled RD. The red boxes are user-marked POIs. }
    \label{fig:cam_view_path}
\end{figure}

Three areas in NYC were of particular interest to our domain collaborators for analysis: the road and open space for a potential evacuation point next to the eastern coast of Manhattan, the Battery park and tunnel area, and road infrastructure at both ends of a bridge on the western end of Manhattan. 
For the POIs marked using red boxes in Fig.~\ref{fig:cam_view_path}, we show the results for our automatic camera viewpoints and path generation for the tiled-display RD in (a) and (b), respectively.
Since the tiled-display is a single-screen setup with a relatively small FoV, we adjust the default weights for the viewpoint optimization such that the POI is positioned towards the center of the display ($\omega_u = 0.9$, $A=\pi/12$) and nearer to the camera ($\omega_d = 0.75$, $K=500m$). 
Conversely, utilizing the RD very large FoV, we adjusted the weights such that the POI could be anywhere within the view of the 4 walls ($\omega_u = 0.1$,  $A=\pi/6$), and enable comparison by allowing multiple POIs to be viewed at an instance ($\omega_d = 0.25$, $K=1000m$).
It took 120s to determine camera views for the single display and 75s for the RD.

\begin{figure}
    \centering
    \includegraphics[width = \linewidth]{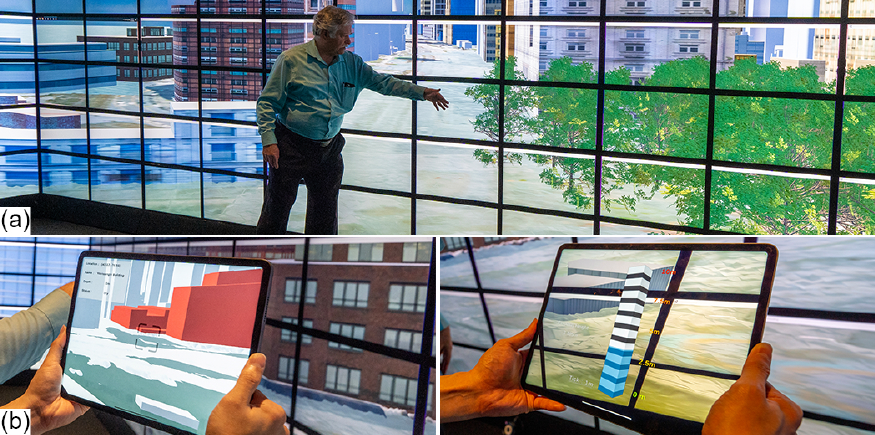}
    \caption{Interaction with data in the RD can be performed either (a) naturally or (b) using our auxiliary (aux) display system.}
    \label{fig:interactions_aux}
\end{figure}

Finally, in Fig.~\ref{fig:interactions_aux}, we demonstrate how users can explore the data by either performing natural pan-and-zoom interactions, as in Fig.~\ref{fig:interactions_aux}(a), or using the aux display  to view additional contextual information, as  in Fig.~\ref{fig:interactions_aux}(b).
Using a game controller, users can manually navigate the scene or use our automatic navigation path to view the marked POIs, and update the timepoint of the flooding scenario.

\subsection{Domain Expert Feedback}
To evaluate Submerse, we conducted two workshops, a specialized one with domain experts and the other with college students without domain knowledge.
The specialized workshop was a follow-up with the focus group we had invited to gather our Submerse system requirements (Sec.~\ref{sec:domain_reqs}).
These included 10 participants: 2 experts from each, NOAA, NWS, NYC Emergency Management, and 2 atmospheric scientists from different academic institutions.
With this group, the primary goal was to evaluate the effectiveness of Submerse in achieving CANVASS aim of facilitating a better understanding of flooding scenarios among stakeholders.
The workshop was divided into two segments. 
The first segment followed NWS protocol of presenting an upcoming storm to the participants using standard NOAA visualizations. 
These included showing a 2D flood layer, with a light-to-dark blue transfer function for flood intensity, on a top-down view of NYC and graphically generated street-level views of five selected POIs (marked by a NOAA expert).
The operator incremented the flood simulation time interval for both visualizations, allowing 30s between each increment. 
After the initial presentation, the experts were allowed to explore the visualizations and data on their own.
The identity of the storm simulation dataset was not revealed to the participants as it would have influenced their perception due to prior knowledge or memory. 
The participants were asked to express and follow the steps they would have normally taken to analyze the provided information.  

\begin{figure}
    \centering
    \includegraphics[width=\linewidth]{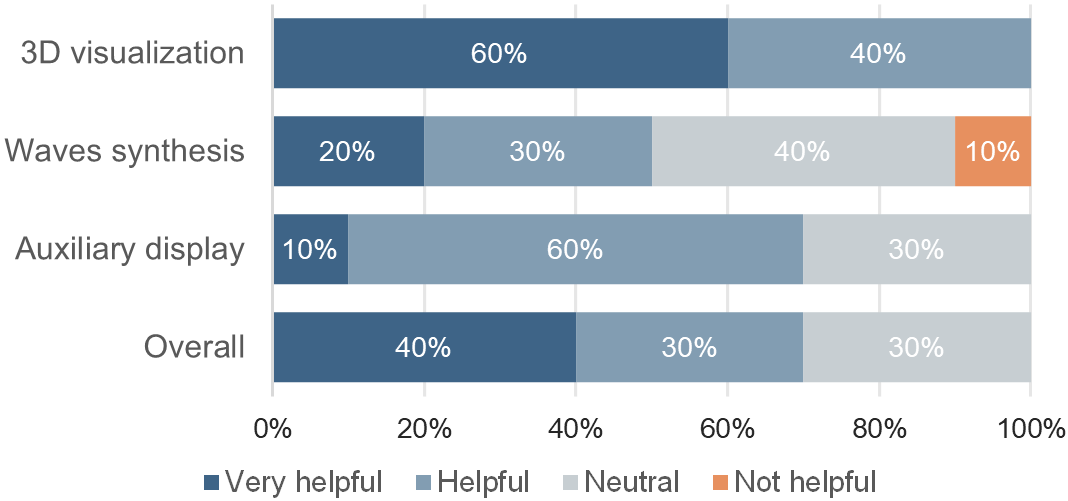}
    \caption{Survey results from our specialized workshop. }
    \label{fig:likert2}
\end{figure}

In the second segment, the experts were given a similar orientation using Submerse in the RD.
We showed our 3D flooding visualization for the automatically generated views using the same input POIs as in the first segment (see Fig.~\ref{fig:cam_view_path}) and incremented the time intervals at the same rate. 
Since the shared view was operator-controlled, we allowed experts to request changing the time intervals for their discussion and analysis.
At the end, we asked the experts to rate Submerse components, namely the 3D visualization, the waves synthesis, the aux display, and the overall system on a Likert scale from \textit{Very helpful}  to \textit{Not helpful}, and asked them for their feedback
(Fig.~\ref{fig:likert2} shows the findings).

\begin{figure}
    \centering
    \includegraphics[width = \linewidth]{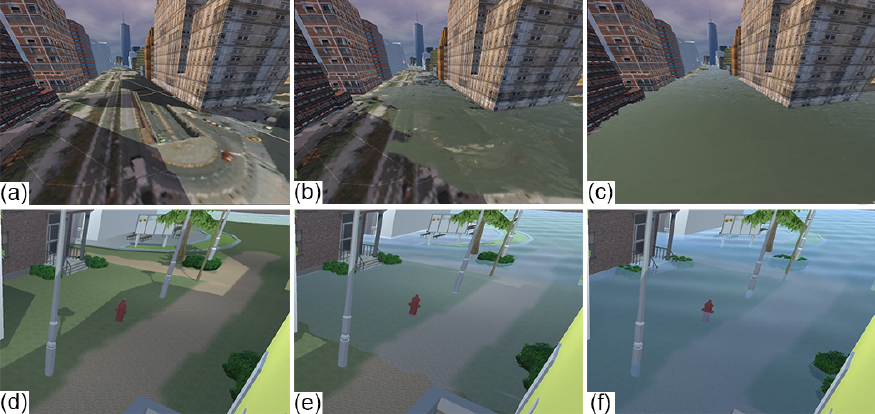}
    \caption{Scenes from the specialized workshop showing NYC in (a)-(c), and the college workshop showing a fictitious scene in (d)-(f). For both, we show no flooding, mid-level, and maximum flooding, respectively.  }
    \label{fig:workshops}
\end{figure}

From the interviews and feedback received, all experts appreciated Submerse. 
One unanimous comment was that the 3D visualization and sense of realism, coupled with immersion, positively transformed the way they planned for preparation, evacuation, and resilience measures, in contrast to currently adopted 2D methods.
Specifically, the to-scale rendering aided  in having a clearer spatial perception of the flooding levels with respect to the surroundings, for instance, the buildings, streets, and open areas.
To this end, the participants found most effective what they called the \textit{street-view}, as shown in Fig.~\ref{fig:interactions_aux} and Fig.~\ref{fig:workshops}(a)-(c), as they could visualize the flood level with respect to their own height. 
One participant reflected that  ``the display of flooding in 3D with actual buildings was a much more effective method of communicating risks, rather than a numerical forecast or a 2D risk map."
In comparing our POI camera viewpoints with the NOAA renderings, all 10 participants mentioned that the Submerse was \textit{More Helpful}.
This was also reflected during manual scene interaction. 
In some instances, participants requested to change the camera views to explore more views around the POI.
However, after multiple camera transformations using a game controller, they requested to return back to the original Submerse view.
Qualitatively, the participants appreciated that the POI was positioned in the scene with respect to the flooding direction.   
While transitioning between POIs using our automatic navigation, one participant commented that the visualizations in the RD felt like ``a helicopter ride during a catastrophe where [they] get a wholesome view of the city, and then go down at street level to investigate its intensity."
Two participants, however, experienced VR sickness during the navigation and one suggested that they would have preferred an instant scene change from one POI to the next.

Overall, the shared experience exploring the data allowed for greater insights as participants were able to instantaneously take into account the diversity of domain expertise for their analysis and decisions.
An example of this was witnessed when an atmospheric scientist explained to an emergency manager how an incoming storm surge would result in a drastic increase in flooding levels of an area and thus may be considered a priority in evacuation plans.
The aux display was welcomed by the participants, especially as they could now make quantitative measurements and seamlessly integrate their domain data visualization for personal-level analysis as well as share their findings collaboratively.
\textit{Occlusion removal} was often utilized to get a clearer view of the surrounding landscape and road network.
Finally, the participants greatly acknowledged wave synthesis for flow direction.  
Results from Fig.~\ref{fig:workshops}(a)-(c), the Battery Tunnel area, were of a particular learning experience.
For each incremental timepoint, the participants were able to observe, for the first time, how tidal surges from different directions contributed to increasing area flooding levels.  
This information was not perceivable otherwise using a still water surface mesh rendering.

The participants that rated the overall Submerse system as \textit{Neutral} had two similar comments: firstly, while the 3D visualizations help better appreciate the flooding impact, the planar top-down view is still an effective method to see the region-wide impact, and secondly, the dependency on the RD means that this system is not easily accessible. 
For the former, we plan to extensively broaden the visualization application and interaction methods of the aux display system in the future, which would also incorporate the standard visualizations accustomed by the experts. 
For the latter, we also plan to extend Submerse to VR HMD in the future. 

A separate workshop with 80 college students was conducted to evaluate the impact of standard weather forecasting graphics compared to Submerse visualizations.
Following a pre-recorded narration of a storm weather forecast, the participants were presented with a hurricane flooding scenario for a fictitious college campus.
We deployed Submerse on a single-screen desktop setting, without an aux display. 
Moreover, the visualizations were designed by an undergraduate visualization student with feedback from an atmospheric scientist and a journalist. 
Thus, supporting Submerse aim of being a modular framework for different display paradigms.
In terms of taking action in favor of preparedness, participants agreed that the visualizations aided their decision-making. 
A general consensus was that the ability to ``see" the flooding, which would otherwise be narrated using words, such as \textit{``we are expecting 3-5 feet of storm surge flooding,"} or 2D graphs, allowed them to understand the forecast severity and  made them more likely to avoid that situation. 
Further details of the workshop procedure and its findings can be found in our  paper~\cite{colle2023risk}.

\section{Conclusion and Future Work}
We have presented Submerse, an application for visualizing 3D flooding scenarios rendered using numerical simulation and GIS data, along with a scalable framework for deploying the application in custom immersive display systems. 
For rendering the flooding level, we have developed a real-time approach to construct a smooth surface mesh, coupled with animated tidal waves to depict flooding direction.
We achieve this by constructing a dynamic quadtree that interactively updates based on the camera view frustum. 

Specifically for immersive visual systems, we have introduced two novel interaction modalities: an automatic camera view-finding and a path-generation algorithm, and an AR-based aux display system. 
We find optimal camera views for marked POIs based on the screen layout and viewing parameters of the POI with respect to the layout. 
For traversing through the views, our path-generation algorithm is designed to avoid collision and reduce VR-sickness, while also facilitating effective scene exploration along the path. 
The aux display is a unique design that compliments the visualizations on the shared screen space by displaying additional contextual information about the virtual scene, using AR. 
The resulting visualizations and particularly the utility of an immersive display system have been positively evaluated by domain experts and other stakeholders.
While Submerse is designed for flood visualization, the interaction techniques can be generalized for other VR applications.
We aim to investigate in the future complex navigation paths that allow \textit{street level} fly-throughs and develop interaction methods to extend the aux system as a focus+context tool for scientific/information visualization.

Based on feedback received during evaluation, we identified a limitation that while Submerse enhances a sense of realism, it does not fully achieve the features of a decision system. 
Therefore, for future work, we plan to integrate tools that will allow stakeholders to visualize outcomes by interactively applying scenarios for mitigation and evacuation measures. 
Moreover, one natural limitation of large immersive facilities is their immobile nature.
To this end, we plan to extend the Submerse system to support VR HMDs and include a framework for co-located as well as remote collaboration across all devices.
Finally, we aim to conduct further research to extend the Submerse visualization and system framework for stereo display systems.

\vspace{-1em}
\ifCLASSOPTIONcompsoc
  \section*{Acknowledgments}
\else
  \section*{Acknowledgment}
\fi

This project was supported in part by NSF grants OAC1919752, ICER1940302, and IIS2107224.

\ifCLASSOPTIONcaptionsoff
  \newpage
\fi



%



\bibliographystyle{IEEEtran}
\bibliography{IEEEabrv,refs}

%

\begin{IEEEbiography}[{\includegraphics[height=1.25in,width=1in,keepaspectratio]{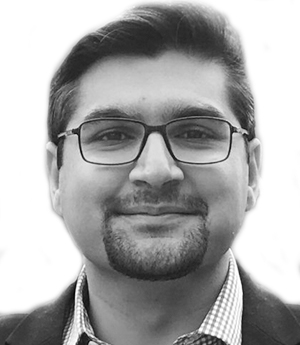}}]{Saeed Boorboor} is a Principal Research Scientist at the Center for Visual Computing, Stony Brook University. He received his Ph.D. in  Computer Science from Stony Brook University and Bachelor's degree in Computer Science from School of Science and Engineering, Lahore University of Management Sciences, Pakistan. His research interests include scientific visualization, immersive analytics, and medical imaging.
\end{IEEEbiography}

\begin{IEEEbiography}[{\includegraphics[height=1.25in,width=1in,keepaspectratio]{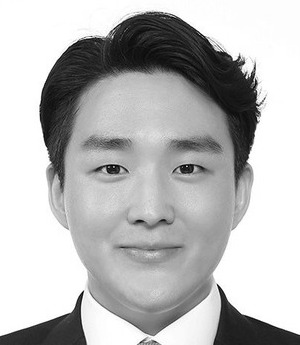}}]{Yoonsang Kim} is currently pursuing a Ph.D. degree in  Computer Science at Stony Brook University. He received his Bachelor's degree in Computer Science from Soongsil University, South Korea, and Master's degree in Computer Science from Stony Brook University. His research interests include augmented reality, virtual reality, data visualization, computer graphics, and human-computer interaction.
\end{IEEEbiography}


\begin{IEEEbiography}[{\includegraphics[height=1.25in,width=1in,keepaspectratio]{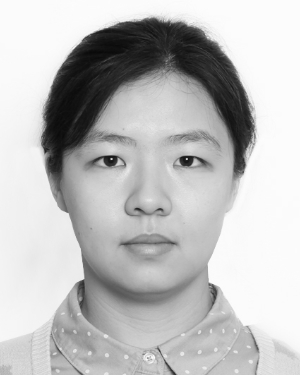}}]{Ping Hu} is currently pursuing a Ph.D. degree in Computer Science at Stony Brook University.  She received her Bachelor's degree in Physics at Shandong University, China. Her research focuses on computational camera control in scientific visualization and computer graphics. 
\end{IEEEbiography}

\begin{IEEEbiography}
[{\includegraphics[height=1.25in,width=1in,keepaspectratio]{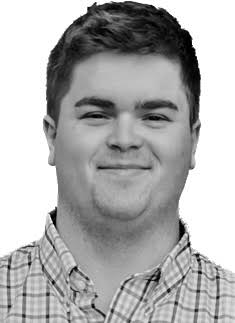}}]{Josef Moses} is currently pursuing a Master degree in Atmospheric Science at Stony Brook University. His research interests include meteorology, risk communication, severe weather communication, and social science.
\end{IEEEbiography}

\begin{IEEEbiography}
[{\includegraphics[height=1.05in,keepaspectratio]{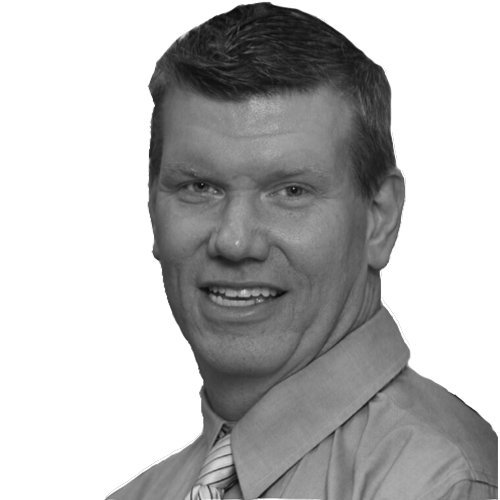}}]{Brian A. Colle}
is a Professor and Division Head for Atmospheric Sciences (since 2018) in the School of Marine and Atmospheric Sciences at Stony Brook University. He has published over 125 papers during the last 25 years on various aspects of extreme weather and climate, including the application and improvement of weather prediction models for these events. He is a Fellow of the American Meteorological Society. He received his Ph.D. in Atmospheric Sciences from the University of Washington (1997).
\end{IEEEbiography}

\begin{IEEEbiography}[{\includegraphics[height=1.25in,width=1in,keepaspectratio]{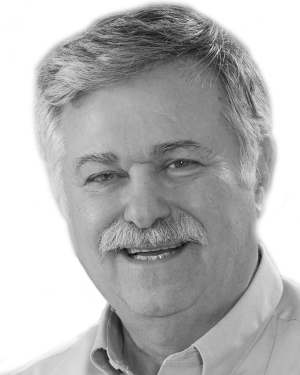}}]{Arie E. Kaufman} is a Distinguished Professor of Computer Science, Director of Center of Visual Computing, and Chief Scientist of Center of Excellence in Wireless and Information Technology at Stony Brook University. He served as Chair of Computer Science Department, 1999-2017. He has conducted research for $>$40 years in visualization, VR and graphics and their applications, and published $>$350 refereed papers. He was the founding Editor-in-Chief of IEEE TVCG, 1995-98. He is an IEEE Fellow, ACM Fellow, National Academy of Inventors Fellow, recipient of IEEE Visualization Career Award (2005), and inducted into Long Island Technology Hall of Fame (2013) and IEEE Visualization Academy (2019). He received his Ph.D.in Computer Science from Ben-Gurion University, Israel (1977).
\end{IEEEbiography}




\end{document}